# 4D Seismic History Matching Incorporating Unsupervised Learning

Clement Etienam[1]

[1] School of Applied Mathematics, the University of Manchester, Manchester, M13 9PL, UK

## Abstract

The work discussed and presented in this paper focuses on the history matching of reservoirs by integrating 4D seismic data into the inversion process using machine learning techniques. A new integrated scheme for the reconstruction of petrophysical properties with a modified Ensemble Smoother with Multiple Data Assimilation (ES-MDA) in a synthetic reservoir is proposed. The permeability field inside the reservoir is parametrised with an unsupervised learning approach, namely K-means with Singular Value Decomposition (K-SVD). This is combined with the Orthogonal Matching Pursuit (OMP) technique which is very typical for sparsity promoting regularisation schemes. Moreover, seismic attributes, in particular, acoustic impedance, are parametrised with the Discrete Cosine Transform (DCT). This novel combination of techniques from machine learning, sparsity regularisation, seismic imaging and history matching aims to address the ill-posedness of the inversion of historical production data efficiently using ES-MDA. In the numerical experiments provided, I demonstrate that these sparse representations of the petrophysical properties and the seismic attributes enables to obtain better production data matches to the true production data and to quantify the propagating waterfront better compared to more traditional methods that do not use comparable parametrisation techniques.

## 1. Introduction

It is well-known that time-lapse seismic data can indicate changes in the rock elastic properties caused by reservoir depletion when the water-front evolves inside a producing reservoir. Calibration of subsurface structures is an essential step in the modelling of fluid flow processes in a plethora of geo-environmental applications such as petroleum and geothermal reservoir monitoring. History matching is known to be an ill-posed inverse process. Its primary goal is to find suitable reservoir model parameters honouring observations (and possibly other constraints) by integration of dynamic data (e.g., gas and oil rate, bottom hole pressure, water cut and subsidence/uplift) and static data (e.g., core, logging, and seismic) (Aanonsen et al., 2009). Ensemble-based data assimilation methods have been used for Seismic History Matching (SHM) for calibrating subsurface models that are tuned and conditioned to dynamic observations (Luo et al., 2018).

Seismic is a vital tool used for reservoir monitoring, exploration and characterisation and management. When compared to dynamic production data which is normally utilised in history matching, seismic data is obtained in lesser frequency during production time but has a much denser sample space. Hence, seismic data can be used to provide vital information for reservoir model calibration to complement production data (Luo et al. 2018).

In recently published work, the ensemble Kalman filter (EnKF) (Evensen 1994, Law et al 2012, Stuart 2010), the Ensemble Smoother (ES) (van Leeuwen & Evensen 1996) and Ensemble Smoother with Multiple Data Assimilation (ES-MDA) (Emerick & Reynolds 2013), as well as iterative variants of ES (Chen & Oliver 2011, Iglesias 2013), have been investigated extensively for history matching/data assimilation problems. However, it has been observed that standard ensemble-based data assimilation approaches have limited capability in preserving non-Gaussian distributions of model parameters such as facies (Aanonsen et al.,2009, Villegas et al., 2018, Lorentzen et al., 2013). Instead, in these classical ensemble-based data assimilation techniques model parameters tend to lose any initially assumed non-Gaussianity which might have been encoded in the starting distributions based on structural constraints observed in hard data. This affects that parameters become increasingly Gaussian during these evolution schemes (Evensen & Eikrem 2018, Kim et al. 2018).

Several approaches have been recently proposed in the literature to maintain structural non-Gaussian features in these techniques. For example, Zhou et al. (2012) use normal score transforms in the ensemble-based data assimilation techniques to preserve non-Gaussian features of model parameters. In that approach, non-Gaussian model parameters are converted first into Gaussian model-states parameters using a normal score transform. EnKF is then applied to these Gaussian parameters. When completed, the final Gaussian profiles are transformed back to the corresponding non-Gaussian parameter representations by back-transformation. In this logic, transformation can take advantage of parameterisation for the cases where the number of required transformed parameters is smaller than the number of original parameters, which helps to reduce computational complexity and to emphasise certain dominant features in the parameters.

A particularly popular model used in such parametrisation techniques is the discrete cosine transform (DCT), which has been employed in the framework of reservoir characterisation by (Jafarpour & McLaughlin, 2007) (Liu & Jafarpour, 2013). An alternative approach is provided by the level set technique (Moreno & Aanonsen 2007, Dorn & Villegas 2008, Villegas et al. 2018, Chang et al. 2010, Lorentzen et al., 2013) which is particularly useful for the modelling of reservoirs where sharp interfaces between different geological regions are present. Moreover, sparse geologic dictionaries (Etienam et al., 2018, Kim et al., 2018, Khaninezhad et al., 2012) including Fourier transform-based methods and specific DCT based techniques are capable of capturing essential features such as patterns and main shapes of a facies channel reservoir (Khaninezhad, et al., 2012). However, they often reveal a deficiency in describing a clear contrast among different facies because of potential data loss during the corresponding inverse transformations (Kim et al. 2018, Khaninezhad et al., 2012, Tarrahi & Afra, 2016).

In the current paper, I will employ a sparse coding approach which is closely related to parametrisation techniques described above. In this approach, sparsity related concepts are applied to both, petrophysical (in particular permeability) and seismic parameters. Sparse coding entails the process of constructing representation coefficients based on the target signal and suitably chosen dictionaries (Candes, et al., 2006). These dictionaries contain groups of features that are capable of compactly representing various pre-specified phenomena typically observed in the environment. Accordingly, in geological modelling, sparse geologic dictionaries are utilised to represent spatial models with a sparse linear combination of basis vectors, which constitute essential geologic features of a reservoir (Khaninezhad, et al., 2012). Such sparse dictionaries can then be employed inside an ensemble-based history matching reducing overall computational complexity whereby maintaining structural consistency with prior information. However, obtaining suitable dictionaries for such a purpose is not straightforward. Aharon et al. (2006) demonstrated the efficacy of an unsupervised learning approach, in particular, K-singular value decomposition (K-SVD), for this purpose. Later Sana et al. (2016) built an archive of essential geologic features (called sparse geologic dictionary) from several static reservoir models using K-SVD and used it for calibrating reservoir models with these dictionaries and EnKF.

A similar approach has been taken by Etienam et al. (2018) for calibrating a synthetic channelised reservoir, however using the ES-MDA instead as the data assimilation technique. In this work, the authors construct sparse geologic dictionaries of the permeability field using K-SVD. They incorporate a greedy algorithm, in particular, orthogonal matching pursuit (OMP), for the sparse representations of the static properties (permeability field) which then is updated by ES-MDA. In a similar but slightly different approach, Kim et al. (2018) couple an iterative form of sparse coding in a transformed space with ES-MDA to construct a set of geologically plausible models that preserve the non-Gaussian distribution of lithofacies in a synthetic channelised reservoir. DCT of sand-shale facies is then followed by repeated K-SVD steps to construct sparse geologic dictionaries. This preserves geologic features of the channelised reservoir such as pattern and continuity.

Let me point out that none of the above mentioned two approaches has integrated seismic attributes into the inverse problem, in contrast to the study presented here. I propose a hybridised ES-MDA algorithm that implements sparse coding to outperform previous seismic history-matching methods in particular for highly non-Gaussian model parameters. The fundamental contribution of this paper is the simultaneous use of DCT analysis for the chosen seismic data combined with K-SVD and OMP based sparse analysis for petrophysical model parameters, and both integrated into an iterative ensemble history matching algorithm, here the ES-MDA. The main ingredient of the idea proposed in this paper SHM framework resides in two aspects. First I reduce the size of big data (which is the acoustic impedance) while retaining sufficient information required for data assimilation; second I parameterise the permeability field with the over-complete dictionary constructed using the unsupervised learning algorithm K-SVD.

The remaining sections of this paper are arranged as follows. Section 2 first introduces the various methodologies being used, in particular, the forward problem, ES-MDA, K-SVD, OMP and DCT. Then the novel SHM technique incorporating unsupervised learning is described. Section 3 presents and discusses the numerical experiments for evaluating this new technique. Section 4 provides some conclusions and indicates possible directions for future work.

## 2. Methodology

This section aims to discuss the various methods that would be used to model the novel SHM technique. Aspects of the forward and inverse problem being used, sparse coding, 4D seismic and DCT for domain compression of the seismic attributes will all be explained in this section.

### 2.1 Forward Problem

For a three-phase flow model of gas, oil and water the forward problem is entirely derived using the combination of Darcy's law and continuity equation. In the following, the index notation $\gamma \, \epsilon \, \{o, w, g\}$ is used for indicating the quantities relating to oil, water and gas respectively. The continuity equation is then written as (Craft & Hawkins 1991),

$$\nabla . \left( \frac{\rho_w}{B_w} u_w \right) - Q_w = \ -\frac{\partial}{\partial t} \left( \frac{\varphi \rho_w}{B_w} S_w \right)$$

Eqn 1

$$\nabla . \left( \frac{\rho_o}{B_o} u_o \right) - Q_o = \ -\frac{\partial}{\partial t} \left( \frac{\varphi \rho_o}{B_o} S_o \right)$$

Eqn 2

$$\nabla . \left( \frac{\rho_g}{B_g} u_g + \frac{R_{so} \rho_g}{B_o} u_o \right) - Q_g = \ -\frac{\partial}{\partial t} \left[ \varphi \left( \frac{\rho_g}{B_g} S_g + \frac{R_{so} \rho_g}{B_o} S_o \right) \right]$$

Eqn 3

where $u_\gamma$ is the Darcy velocity of each phase

$$u_\gamma = -\frac{k_{r\gamma}}{\mu_\gamma} K(\nabla p_\gamma - \rho_\gamma \hat{g} \nabla z)$$

Eqn 4

$$S_o + S_w + S_g = 1$$

Eqn 5

$$p_o - p_w = p_{cow}$$

Eqn 6

$$p_g - p_o = p_{cgo}$$

Eqn 7

Here, $\varphi$ is the effective porosity , $K$ is the absolute permeability, $\hat{g}\nabla z$ is the acceleration due to gravity. $Q_\gamma$ are the source and sink terms representing the phase production and injection, $S_\gamma$ and $p_\gamma$ are the phase saturation and phase pressure, $k_{r\gamma}$ is the phase relative permeability, $\rho_\gamma$ is the phase density, $\mu_\gamma$ is the phase viscosity. $B_\gamma$ is the phase formation volume factor, $R_{so}$ the solution gas-oil ratio, $p_{cgo}$ the gas-oil capillary pressure. $z$ is the vertical coordinate (height) and $\hat{g}$ denotes the magnitude of the gravitational acceleration. Next, I will discuss the ES-MDA algorithm which will be used in solving the history matching inverse problem.

## 2.2 ES-MDA

The history matching problem considered here can be formulated mathematically as

$$min\ J(\boldsymbol{m})$$

Eqn.8

Where $J$ denotes the objective function of the history matching problem and $\boldsymbol{m}$ denotes the state vector composed of reservoir variables (e.g., permeability and facies) to be estimated. The typical expression of $J(\boldsymbol{m})$ for ensemble-based history matching problems is presented as (Emerick & Reynolds, 2013, Oliver et al 2008, Tarantola 2005)

$$J(\boldsymbol{m}) = (\boldsymbol{m} - \boldsymbol{m}^b)^T \boldsymbol{B}^{-1}(\boldsymbol{m} - \boldsymbol{m}^b) + (d^{obs} - d)^T \boldsymbol{R}^{-1}(d^{obs} - d)$$

Eqn 9

In this notation $\boldsymbol{m}^b$ denotes a known estimate for the state vector and the superscript $b$ indicates the background (sometimes denoted prior) distribution; $\boldsymbol{B}$ denotes the covariance matrix of $\boldsymbol{m}^b$; $d^{obs}$ denotes the observed responses; $\boldsymbol{d} = G(\boldsymbol{m})$ is the dynamic vector composed of simulated responses constructed by running a reservoir simulator $G$ for the state vector $\boldsymbol{m}$; and $\boldsymbol{R}$ denotes the covariance matrix of observation error. The right-hand side of Eqn.9 is the addition of background and observation error terms (Emerick & Reynolds, 2013). $\boldsymbol{m}$ in principle can contain any unknown variables such as facies indexes, permeability/porosity field, coefficients of discrete cosine functions or sparse coefficients.

$\partial J(\boldsymbol{m})/\partial \boldsymbol{m} = 0$ can be used to derive the minimum of the cost function and update equation for $\boldsymbol{m}$ as (Emerick and Reynolds 2013, Nocedal 1999, Sherman & Morrison 1950)

$$\boldsymbol{m_i} = m_i^b + \boldsymbol{C_{md}}\big(\boldsymbol{C_{dd}} + \alpha_p C_d\big)^{-1}\big(d_i^{pert} - \boldsymbol{d_i}\big) \, for \, i = 1 \ldots N_{ens}$$

<div align="right">Eqn 10</div>

where the subscript $i$ denotes the $i^{th}$ ensemble member; $\boldsymbol{C_{md}}$ denotes the cross-covariance matrix of $\boldsymbol{m}$ and $\boldsymbol{d}$; $\boldsymbol{C_{dd}}$ denotes the auto-covariance matrix of $\boldsymbol{d}$; $\alpha_p$ is the coefficient to inflate $C_d$, which denotes the covariance matrix of the observed data measurement error (Emerick and Reynolds, 2013); $d^{pert}$ denotes the observation data perturbed by the inflated observed data measurement error; and $N_{ens}$ is the ensemble size (i.e., number of reservoir models as columns of the matrix in the ensemble). Conventionally, ensemble-based history matching updates $N_{ens}$ reservoir models simultaneously. In Eqn.10, $\boldsymbol{C_{md}}\big(\boldsymbol{C_{dd}} + \alpha_p C_d\big)^{-1}$ denotes the Kalman gain $K$, which is normally computed with regularization by SVD using 99.9% of the total energy in singular values (Emerick and Reynolds, 2013, Oliver et al 2008, Law et al 2012, Hanke 1997).

ES-MDA assimilates every state vector $N_a$ times using an inflated covariance matrix of measurement error (Emerick and Reynolds, 2013). In this case, $N_a$ is the number of assimilations in ES-MDA. $\boldsymbol{C_{md}}$ and $\boldsymbol{C_{dd}}$ are defined as follows:

$$\boldsymbol{C_{md}} = \frac{1}{N_{ens} - 1} \sum_{i=1}^{N_{ens}} (\boldsymbol{m_i} - \boldsymbol{\bar{m}})\big(\boldsymbol{d_i} - \boldsymbol{\bar{d}}\big)^T$$

<div align="right">Eqn 11</div>

$$\boldsymbol{C_{dd}} = \frac{1}{N_{ens} - 1} \sum_{i=1}^{N_{ens}} \big(\boldsymbol{d_i} - \boldsymbol{\bar{d}}\big)\big(\boldsymbol{d_i} - \boldsymbol{\bar{d}}\big)^T$$

<div align="right">Eqn 12</div>

where $\boldsymbol{\bar{m}}$ denotes the mean of state vectors and $\boldsymbol{\bar{d}}$ denotes the mean of dynamic vectors.

In ES-MDA, $\alpha_p$ is normally constrained to

$$\sum_{p=1}^{N_a} \frac{1}{\alpha_p} = 1$$

<div align="right">Eqn 13</div>

$$d_i^{pert} = d^{obs} + \sqrt{\alpha_p} C_d^{\frac{1}{2}} z_{d,i} \, for \, i = 1 \ldots N_{ens}$$

<div align="right">Eqn 14</div>

The second quantity on the right-hand side of Eqn 14 is known as the perturbation term. It reflects the uncertainty associated with data measurement and processing. The stochastic characteristics of $C_d$ are reflected by $z_d \sim N\big(0, I_{N_d}\big)$. $z_d$ is the random error matrix to observations, which is constructed with a mean of zero and a standard deviation of $I_{N_d}$, where $N_d$ is the number of time steps found in the observations. For more details see (Emerick and Reynolds 2013).

The goal in this approach is to use sparse representations inside the ES-MDA workflow. Sparsity, concerning petrophysical parameters requires efficient tools which will be described in the next paragraph. In particular, I will explain the unsupervised learning algorithm K-SVD and sparse coding with OMP.

## 2.3 Sparse Coding

Sparse coding can be described as the process of computing the representation coefficients, X, based on the given signal $Y$ and a dictionary $D$. This process, commonly referred to as atom "decomposition", is proven to be an NP-hard problem (Aharon, et al., 2006), which is very difficult to solve. Therefore, it is often replaced by a "pursuit algorithm" that finds an approximate solution instead. Among the most straightforward pursuit algorithms are the Matching Pursuit (MP) and Orthogonal Matching Pursuit (OMP) (Tropp & Gilbert, 2007). These are known as greedy algorithms which select the dictionary atoms sequentially. These techniques are straightforward, involving the computation of inner products between the signal and the dictionary atoms, and possibly deploying some least squares solvers or projections. In this work, I employ the OMP algorithm which is described further below.

### 2.3.1 K-SVD Algorithm for obtaining a dictionary

A key criterion of sparse coding is the identification of a basis (domain) in which the field or signal under consideration can be modelled as sparse (Sana et al., 2016) A sparse domain describes one whereby any signal located in that domain can be adequately represented by a linear combination requiring a few basis elements (Candes, et al., 2006). The Wavelets (Jafarpour, 2011) and the Discrete Cosine Transform (DCT) (Jafarpour & McLaughlin, 2007) are examples of such sparse domains. In this work, I construct a sparse domain that is suited for inverse problems in reservoir engineering (history matching). Such algorithms modified for history matching are known as *dictionary learning algorithms* (Elsheikh, et al., 2013).

The term 'unsupervised learning' is a part of machine learning that learns from test data which has not been labelled or classified. Unsupervised learning identifies patterns or clusters in the training data and predicts the trends in each new piece of data (Bishop 2006). With this in mind, one drawback of K-SVD is its large size of sparse geologic dictionaries (Khaninezhad et al., 2012). References from the literature show that sparse coding implementation with a transformation of parameter space could significantly reduce both computational complexity and costs that are simultaneously required for model calibration (Khaninezhad et al., 2012, Etienam et al., 2018, Kim et al., 2018, Sana et al., 2016).

For a set of training signals $Y$ of size $N_Y$ (which is also the dimension of the unknown permeability field) the dictionary learning algorithm finds the basis elements necessary to construct a dictionary $D$ whereby its sparse linear combinations represents each of the vector space found in the training set of signals $Y$. The corresponding weights of the signals are given as $X$ and are calculated by solving the optimisation problem,

$$\{X, D\} = \arg\min_{X,D} \sum_{i=1}^{N_Y} \|Y_i - DX_i\|_2^2 + \gamma \|X_i\|_1$$

Eqn.15

$$\|D_i\|_2^2 \leq 1, \quad \forall i = 1, \dots N_Y$$

Eqn. 16

$X_i$ and $Y_i$ are the individual training signals and coefficient vectors in $X$ and $Y$, respectively. $D_i$ are the atoms of the dictionary $D$ selected such that the $\ell_1 - norm$ of $X_i$ satisfies Eqn 18. Finding $X$ and $D$ requires us to solve a non-convex problem, which is difficult even in an approximate manner (Elsheikh, et al., 2013).

Notice that Eqn 15 amounts of a simultaneous minimisation of $X, D$. A computationally feasible dictionary learning algorithm that can solve this problem is the K-SVD algorithm. The K-SVD is NP-hard (level of computational complexity, i.e. *non-deterministic polynomial time*), and it solves the $\ell_0 - norm$ equivalent of the problem enumerated in equation Eqn 15 by using a cyclic three-step approach (Aharon, et al., 2006) indicated in Figure 1. The first of the three steps selects the cluster mean using K-means clustering. The second step is the sparse coding stage for updating the coefficient of the matrix $X_i$ for a given iterate of $D$ making use of any pursuit algorithm (a sparsity inducing method). Here OMP will be used which is explained in the next paragraph. In the third step, each column of the dictionary elements $D$ will be updated in a sequential order by changing the values of its columns and also updating the sparse coefficients $X_i$. This is done using the signals $Y_i$ such that $Y = DX$, see again Figure 1. For further studies on the K-SVD, the reader may refer to (Aharon et al., 2006).

### 2.3.2 Orthogonal Matching Pursuit

In the process of establishing a suitable over-complete dictionary, we need to be able to represent vectors in a sparse form by atoms of this proposed dictionary. For this purpose, I employ OMP which works as follows. For an estimate of sparsity $T_0$ for a field that is represented by a vector of weights $X$ in a d-dimensional space, the OMP algorithm attempts to recover the signal $Y$ by using a linear combination of $T_0$ basis atoms (basis elements) from such an over-complete dictionary $D$. The OMP algorithm finds such a sparse representation by solving the following optimization problem

$$\min_{\mathcal{X}} \|\mathcal{X}\|_0 \ \ subject \ to \ \|Y - DX\|_2 < \eta$$

<div align="right">Eqn 17</div>

Practically it solves this optimisation problem by projecting the data iteratively onto a dictionary and then finding a set of basis elements that are maximally correlated with the residuals. $\|X\|_0$ is the $\ell_0$-norm and it represents the signal sparsity level of $\mathcal{X}$. $\eta$ is the error tolerance.

The OMP algorithm requires as inputs an estimate of sparsity $T_0$ for the sparse signal $X$, the dictionary $D$ of rank $N_Y \times d$ with $d$ being the same as the size of the row of the sparse coefficients . The idea in OMP is in constructing an approximation to the signal $Y$ (the spatial permeability field) by undergoing an iteration process. During each iteration, the local optimum of the solution is computed. This technique is calculated by finding the column vectors in $D$ (the over-complete K-SVD learned dictionary) which closely matches a residual vector indicated as $r$. The reader may refer to Tropp & Gilbert, (2007) for further information.

As already mentioned before, in this work the K-SVD combined with OMP is used to generate the over-complete dictionary. This is done off-line and once before the commencement of history matching with the ES-MDA. A schematic showing the implementation of the K-SVD is shown in Figure 1.

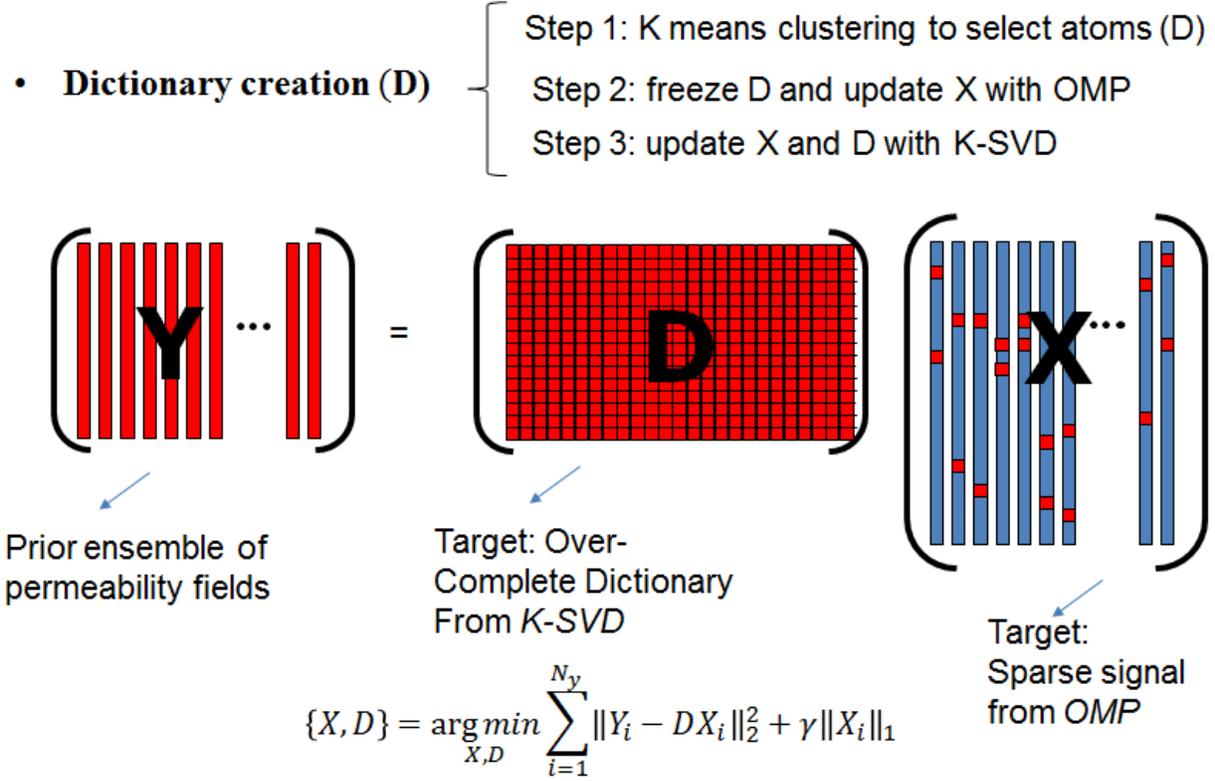

$$\{X, D\} = \underset{X,D}{\arg\min} \sum_{i=1}^{N_\gamma} \|Y_i - DX_i\|_2^2 + \gamma \|X_i\|_1$$

**Figure 1**: Schematic showing the K-SVD algorithm

Next, I will discuss on the petro-elastic model being used in this work and the general 4D seismic approach

### 2.4 4D Seismic

In general, constructing a proper Petro-Elastic Model (PEM) is vital to the implementation and success of SHM I assume that the PEM is perfect and use the soft sand model as the PEM (Luo et al., 2018, Mavko et al., 2009). The PEM assumes that the cement is deposited away from the grain contacts. It also considers that the initial framework of the un-cemented sand rock is a densely random pack of spherically shaped grains having the porosity around 36%, which is the maximum porosity value that the rock could have before the suspension. I denote this parameter as the critical porosity $\varphi_c$ (Luo et al, 2018). The shear modulus $\mu_{HM}$ and dry bulk modulus ($K_{HM}$) at critical porosity is then calculated using the Hertz-Mindlin model as denoted in (Mindlin, 1949, Luo et al, 2018)

$$K_{HM} = \sqrt[n]{\frac{C_p^2(1-\varphi_c)^2 \mu_s^2}{18\Pi^2(1-v_s)^2}} P_{eff}$$

Eqn 18

$$\mu_{HM} = \frac{5-4v_s}{5(2-v_s)} \sqrt[n]{\frac{3C_p^2(1-\varphi_c)^2 \mu_s^2}{2\Pi^2(1-v_s)^2}} P_{eff}$$

Eqn 19

where $P_{eff}, \mu_s, \nu_s$ are Poisson's ratio, grain shear modulus and effective stress, respectively. $C_p$ denotes the coordination number and is the average number of contacts per sphere, $n$ is the degree of the root in this case. Here, the $C_p$ and $n$ are 9 and 3, respectively. The modified Lower Hashin-Shtrikman (MLHS) bound is used to find the effective dry moduli for a porosity value less than $\varphi_c$ (Luo et al, 2018, Mavko et al, 2009). MLHS connects two end points in the elastic modulus-porosity plane. One end point corresponds to zero porosity, the other end point ($K_{HM}$ $\mu_{HM}$), corresponds to critical porosity $\varphi_c$ taking the moduli of the solid phase, i.e. quartz mineral ($K_s, \mu s$). For porosity value $\varphi$ between zero and $\varphi_c$, the lower bound for dry rock effective bulk ($K_d$) and shear ($G_d$) are;

$$K_d = \left[ \frac{\frac{\varphi}{\varphi_c}}{K_{HM} + \frac{4}{3}\mu_{HM}} + \frac{\frac{1-\varphi}{\varphi_c}}{K_s + \frac{4}{3}\mu_{HM}} \right] - \frac{4}{3}\mu_{HM}$$

Eqn 20

$$G_d = \left[ \frac{\frac{\varphi}{\varphi_c}}{\mu_{HM} + \frac{\mu_{HM}}{6}Z} + \frac{\frac{1-\varphi}{\varphi_c}}{\mu_s + \frac{\mu_{HM}}{6}Z} \right] - \frac{\mu_{HM}}{6}Z$$

Eqn 21

$K_s$ is solid/mineral bulk modulus and $Z = (9K_{HM} + 8\mu_{HM})/(K_{HM} + 2\mu_{HM})$ The saturation effect is infused using the Gassmann model (Gassmann,1951). The saturated bulk modulus ($K_{sat}$) and shear modulus ($\mu_{sat}$) are defined as;

$$K_{sat} = K_d + \frac{(1 - \frac{K_d}{K_s})^2}{\frac{\varphi}{K_f} + \frac{1-\varphi}{K_s} - \frac{K_d}{K_s^2}}$$

Eqn 22

$$\mu_{sat} = \mu_d$$

Eqn 23

$K_f$ is the effective fluid bulk modulus. For a gas- oil-water mixture (as is the case in the PUNQ-S3 synthetic model), $K_f$ is given by

$$K_f = \left( \frac{S_w}{K_w} + \frac{S_o}{K_o} + \frac{S_g}{K_g} \right)^{-1}$$

Eqn 24

where $K_w$, $K_o$, $K_g$, $S_w$, $S_o$ and $S_g$ are the bulk modulus of water/brine, bulk modulus of oil, bulk modulus of gas, saturation of water/brine, saturation of oil and saturation of gas, respectively. The saturated density (Gassmann, 1951) can be written as;

$$\rho_{sat} = (1 - \varphi)\rho_m + \varphi S_w \rho_w + \varphi S_o \rho_o + \varphi S_g \rho_g$$

Eqn 25

Here $\rho_{sat}$, $\rho_m$, $\rho_w$, $\rho_o$ and $\rho_g$ are saturated density of rock, mineral density, water/brine density, oil density and gas density respectively.

Using equations 18-25, P- and S-wave velocities can be obtained as (Gassmann, 1951);

$$V_p = \sqrt{\frac{K_{sat} + \frac{4}{3}\mu_{sat}}{\rho_{sat}}}$$

Eqn 26

$$V_s = \sqrt{\frac{\mu_{sat}}{\rho_{sat}}}$$

Eqn 27

Since the volumetric scale of seismic data is large, SHM often can be referred to as a "big data" assimilation problem. Some numerical issues may arise for ensemble-based history matching algorithms when assimilating big data, e.g., ensemble collapse of the ensemble when tracking the true signal and large costs in storing and computing the Kalman gain matrices (Luo et al., 2018, Skjervheim 2007). Also, the history matching may become over-determined with big data (such that there is no solution that could match the observational data exactly) as opposed to the under-determined nature required for most successful implementation of history matching which may affect accuracy of the algorithm results (Luo et al. 2018, Tarantola 2005, Skjervheim 2007). Therefore, data reduction is a necessary approach in big data analytics, as seen in geosciences (Luo et al., 2018). I will use DCT for addressing this task concerning the seismic data.

### 2.5 Discrete Cosine Transform

While OMP is employed as described above for obtaining a sparse representation of the petrophysical parameters, a straightforward DCT will be more appropriate in our case for enforcing a sparse representation of the seismic attributes due to its different physical meaning and scale. Due to the low resolution of seismic in the vertical dimension (as opposed to a very fine resolution in the other two dimensions), it is furthermore sufficient to average seismic parameters along this direction on the scale of the reservoir which allows us to use DCT in two dimensions.

The two-dimensional forward DCT of an input image $u(n_x, n_y)$ for an output image $v(\Theta, \omega)$ can be described as (William & Mitchell 1993);

$$v(\Theta, \omega) = \frac{\alpha(\Theta)\alpha(\omega)}{\sqrt{N_x N_y}} \sum_{m=0}^{N_x-1} \sum_{n=0}^{N_y-1} u(n_x, n_y) \; cos\left[\frac{\pi(2n_x + 1\Theta)}{2N_x}\right] cos\left[\frac{\pi(2n_y + 1)\omega}{2N_y}\right]$$

Eqn 28

$$\alpha(j) \equiv \begin{cases} \frac{1}{\sqrt{2}} & j = 0 \\ 1 & otherwise \end{cases}$$

Where $\Theta = 0, \dots, N_x - 1$ and $\omega = 0, \dots, N_y - 1$.

Similarly, the inverse two-dimensional DCT transform is then derived as

$$u(n_x, n_y) = \frac{2}{\sqrt{N_x N_y}} \sum_{\Theta=0}^{N_x-1} \sum_{\omega=0}^{N_y-1} \alpha(\Theta)\alpha(\omega)v(\theta, \omega) \; cos\left[\frac{\pi(2n_x + 1\Theta)}{2N_x}\right] cos\left[\frac{\pi(2n_y + 1)\omega}{2N_y}\right]$$

<div align="right">Eqn 29</div>

Where $n_x = 0, \ldots, N_x - 1$ and $n_y = 0, \ldots, N_y - 1$

For more information on the DCT, the reader may refer to (William & Mitchell 1993).

## 2.6 Novel SHM technique

In this section, I will state the novel history matching approach abbreviated as **SHM-KED** which stands for **S**eismic **H**istory **M**atching incorporating **K**-SVD with **ES**-MDA and **D**CT. The algorithm for the procedure is given next and explained below;

*Algorithm:* **SHM-KED**

---

1. Generate $K_{global} = 2000$ with *FILTERSIM* (Wang et al., 2006) Compute $D$ using K-SVD Algorithm (off-line and once).
2. Select $N_{ens}$ permeability realisations from $K_{global}$ as an initial ensemble. Input ensemble size, $N_{ens} = 200$,
3. Choose the number of data assimilations, $N_a$ and the coefficients $\alpha_p{}^i$ for $i = 1, 2, \ldots . N_a$
4. For $i = 1 \ to \ N_a$
   a. Run the ensemble from time zero and solve the forward problem (Eqn.1-7)
   b. For each member of the ensemble, perturb the observation vector using $d_i^{pert} = d^{obs} + \sqrt{\alpha_p} C_d^{\frac{1}{2}} z_{d,i} \ for \ i = 1 \ldots N_{ens}$, where $z_d \sim \mathcal{N}\left(0, I_{N_d}\right)$
   c. Obtain impedance for the two-time survey using the PEM model (Eqn 18-27)
   d. Obtain sparse representations of impedance using DCT (Eqn 28)
   e. Obtain sparse representation of permeability sparse using OMP Algorithm with over-complete dictionary $D$
   f. For k=1 to production history
      - IF time step = impedance survey instance
      - Update the ensemble of sparse coefficients using Eqn 10 with simulated impedances and production data at time step

      ELSE
      - Update the ensemble of sparse coefficients using Eqn 10 with production data alone

      End (For)

   End (For)

   End program

---

The initial realisations for this algorithm are generated using the Multiple-Point Statistics (MPS) algorithm. 200 realisations of static reservoir parameters, which are the permeability fields, are generated based on the well data. The changes of the dynamic parameters, pressure and saturation, and the simulated pressure-production data are generated using a three-phase flow simulator to obtain an ensemble of simulated states. The simulated impedance maps associated with the simulate states for one particular time step are calculated using the given petro-elastic model. The simulated changes in seismic attribute, in particular, impedance maps, and the well production data, give rise to an ensemble of simulated measurements. The true or reference measurements are the seismic attributes and the historical production data, in particular, real impedance maps, computed from a prior elastic full-waveform inversion scheme. The DCT algorithm parametrises this seismic attribute by selecting the leading cosine basis of these properties. The K-SVD learned dictionary enforces prior structural information on the permeability field during the ES-MDA inversion step. The updated state vector contains the corrected static and dynamic states of the reservoir after a single iteration of the ES-MDA.

## 3. Numerical Experiments/Results

### 3.1 The computational test case PUNQ-S3

In this section, I will test the novel SHM approach on a synthetic reservoir model. The PUNQ-S3 is a small-size synthetic 3D reservoir engineering model (Floris et al. 2001). The PUNQ-S3 model is made up of 19×28×5 grid blocks, where 1761 are active grid blocks. The grid-blocks have equal 180-meter sides in x- and y-directions. The heights of the grid blocks are varying. Figure 2 shows the top layer of the synthetic reservoir, including six production wells (PRO-1, PRO-4 PRO-5, PRO-11, PRO-12, and PRO-15). The reservoir is bounded by a strong aquifer in the west and north and a sealing fault in the east and south. The is no requirement for an injection well as the aquifer ensures high reservoir pressure. The red zone in Figure 2 below denotes a gas cap. There are two sections for the production schedule. The first is considered a history matching phase, and it lasts for 8 years. This section consists of four years of production, one year of extended well testing and a 3-year shut-in period. The second section comprises of 8.5 years of production. Measurements from all six wells were used during the assimilation period. The measurements includes bottom hole pressures, gas-oil ratios, and water cuts. The shut-in pressure has a noise value of 1 bar whereas flowing pressure has a noise level of 3 bar. The gas-oil ratios have a measurement uncertainty of 10 % before gas breakthrough and 25 % after the gas breakthrough. The water cut has 2 % uncertainty before water breakthrough and 5% after. The noise in the measurements is assumed uncorrelated, and I have used the same noise level generating the observations as when running the algorithm. A commercial reservoir simulator (Schlumberger GeoQuest, 2014) is used for reservoir simulation. Facies type is observed at well locations.

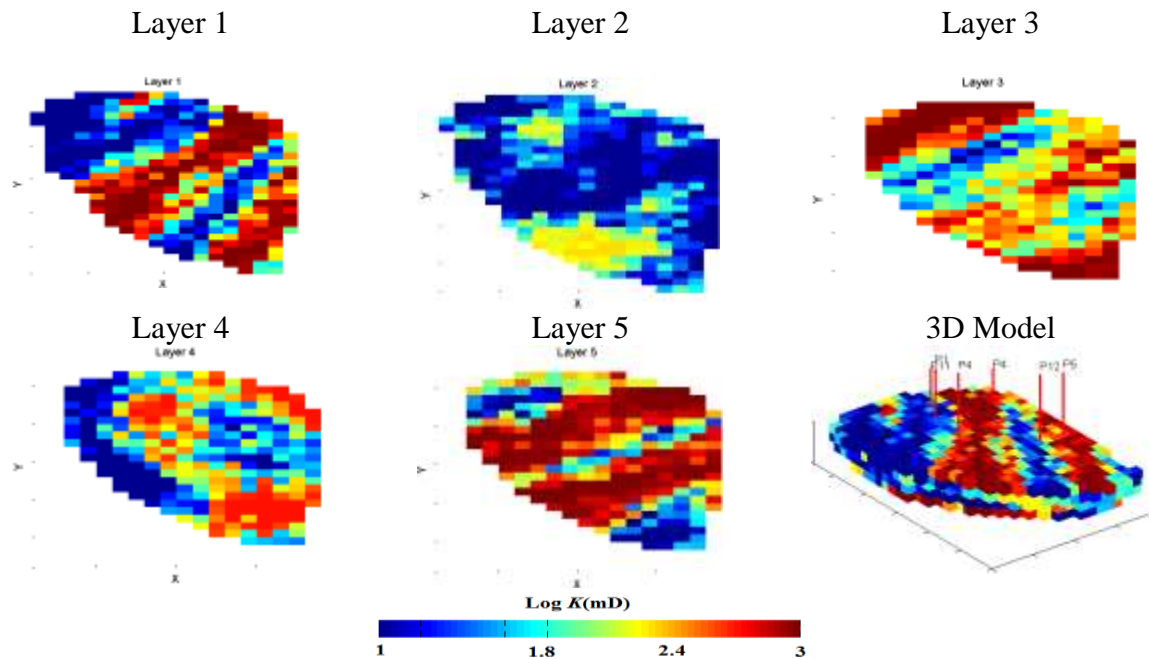

**Figure 2**: permeability field (Log K) of the true PUNQ-S3 synthetic model showing the 2D planar permeability field and 3D model showing the six well locations.

### 3.2 Computer specification

The computer specification for the desktop computer used for conducting numerical experiments is shown in Table 1.

**Table 1**: Computer specification for running the numerical experiments

| Type | Desktop |
|---|---|
| **Model** | Dell Optiplex 790 |
| **Operating system** | Windows 7 Enterprise 64-bit |
| **Processor** | Intel (R) Core(TM) i5-2400 |
| | CPU@ 3.10 GHz,3101 Mhz, |
| | 4Cores(s), 4 Logical |
| | Processor(s) |
| **RAM** | 8.00GB |

Table 1 shows the computer specification used for running the numerical experiments in this paper.

### 3.3 Performance measure RMSE

To quantify our production data match to the true observed data, I use the root-mean-square-error (RMSE) function for each ensemble member ($i$), which is defined as

$$\text{RMSE}(i) = \left( \frac{1}{N} \sum_{k=1}^{N} \sum_{j=1}^{N_{data}^{k}} \left( \frac{D_{obs}^{j}(k) - D_{sim,i}^{j}(k)}{\sigma_{n,j}} \right)^2 \right)^{\frac{1}{2}}$$

Eqn 30

$N$: Number of data assimilation time steps where measurements are assimilated (measurement times)
$N_{data}^{k}$: Number of data collected at each time step $k$
$i$: Ensemble member index
$k$: Time index
$j$: Metric or response (history matched metric or response)
$D_{obs}^{j}(k)$: Observed data metric for metrics $j$ (Data equivalent in state space ensemble) at time step $k$.
$D_{sim}^{j}(k)$: Simulated data from simulator for metrics $j$ (Data equivalent in state space ensemble) at time step $k$.
$\sigma_{n,j}$: Observed data standard deviation for metrics $j$ (Data equivalent in state space ensemble).

### 3.4 Presentation and Discussion of Results

In figure 3, I generate the initial ensemble of horizontal permeability field ($K_x$) using *FILTERSIM* (Wu et al, 2006). To generate the initial ensemble of porosity and permeability (in the z-direction) , I use the following statistical relationship derived from cores of the true synthetic model.

***For Layer 1***

$$\varphi = 0.040228 \times \ln(K_x) - 0.03101$$
$$K_z = 0.88227 \times \ln(K_x) - 0.29112$$

Eqn 31(a)

**For Layer 2**

$$\varphi = 0.022608 \times \ln(K_x) - 0.0066038$$
$$K_z = 0.89976 \times \ln(K_x) - 1.1289$$

Eqn 31(b)

**For Layer 3**

$$\varphi = 0.046974 \times \ln(K_x) - 0.072764$$
$$K_z = 0.69049 \times \ln(K_x) + 1.0074$$

Eqn 31(c)

**For Layer 4**

$$\varphi = 0.025312 \times \ln(K_x) + 0.01088$$
$$K_z = 0.82778 \times \ln(K_x) - 0.56077$$

Eqn 31(d)

**For Layer 5**

$$\varphi = 0.039746 \times \ln(K_x) - 0.038238$$
$$K_z = 0.88227 \times \ln(K_x) - 0.29112$$

Eqn 31(e)

From this the state variable for ES-MDA or SHM-KED becomes;

$$\begin{pmatrix} K_x \\ BHP \\ GOR \\ WCT \end{pmatrix}$$

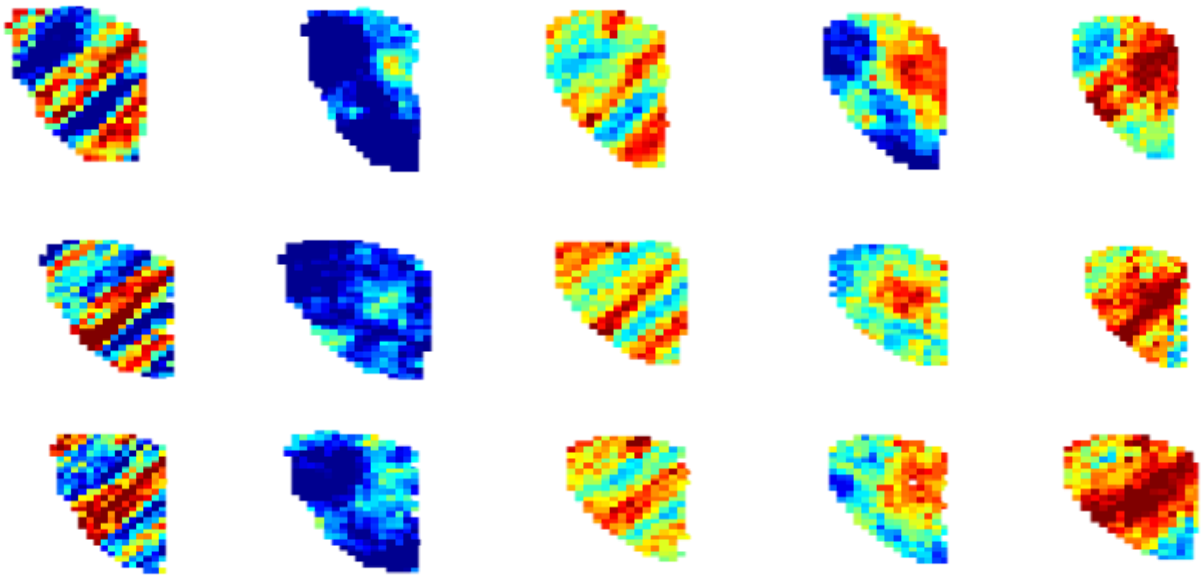

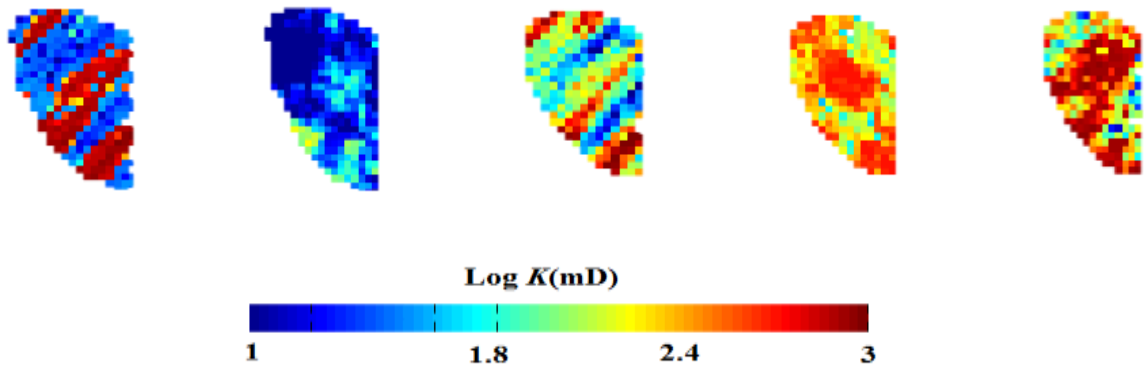

**Figure 3**: an initial model of permeability realisations #21, #11, #31 and #67 generated with the *FILTERSIM* algorithm

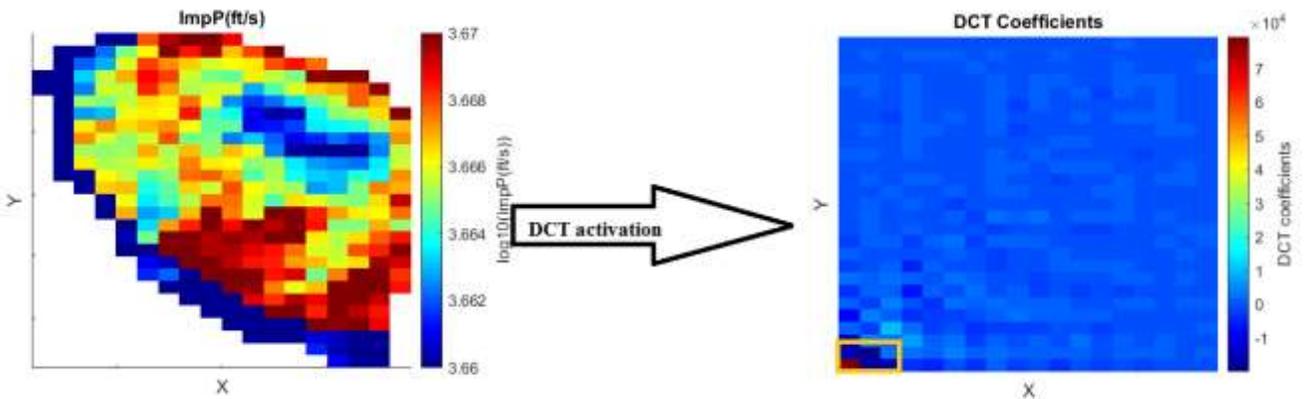

**Figure 4**: Schematic showing how DCT can be used for image compression.

Figure 4 shows how DCT can be used to generate a sparse representation of the true impedance field. The leading coefficient (at the bottom lower corner) represents more than 98% of the information in the original impedance field.

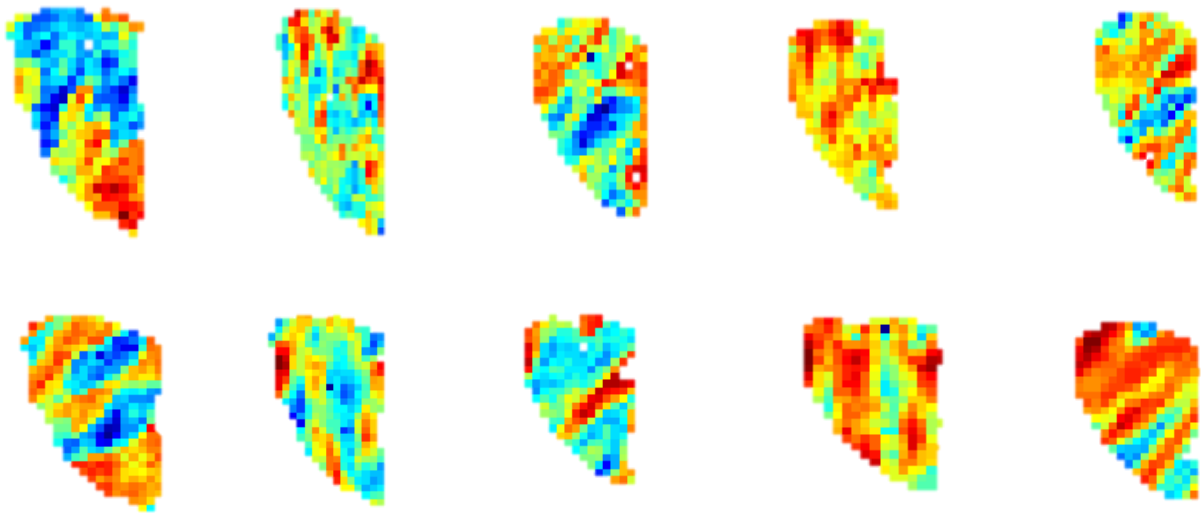

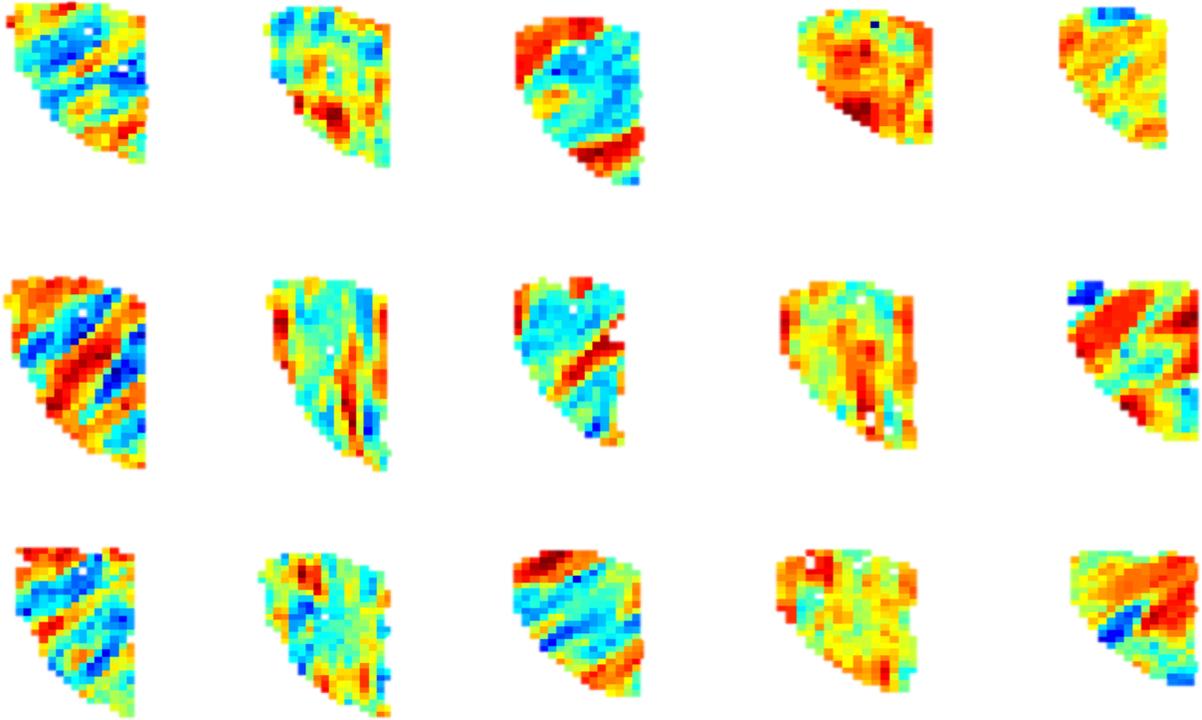

**Figure 5**: Some initial permeability K-SVD basis atoms recovered after dictionary learning

Figure 5 shows some K-SVD basis recovered after activating K-SVD on an initial pool of realisations. Notice the patterns being maintained in this dictionary basis. With this idea, I will enforce prior structural information on the permeability updates during ES-MDA. In this numerical experiment, I use constantly $\alpha_p = 8$ for. $p = 1. N_a$ where $N_a = 8$ for the ES-MDA data assimilation algorithm.

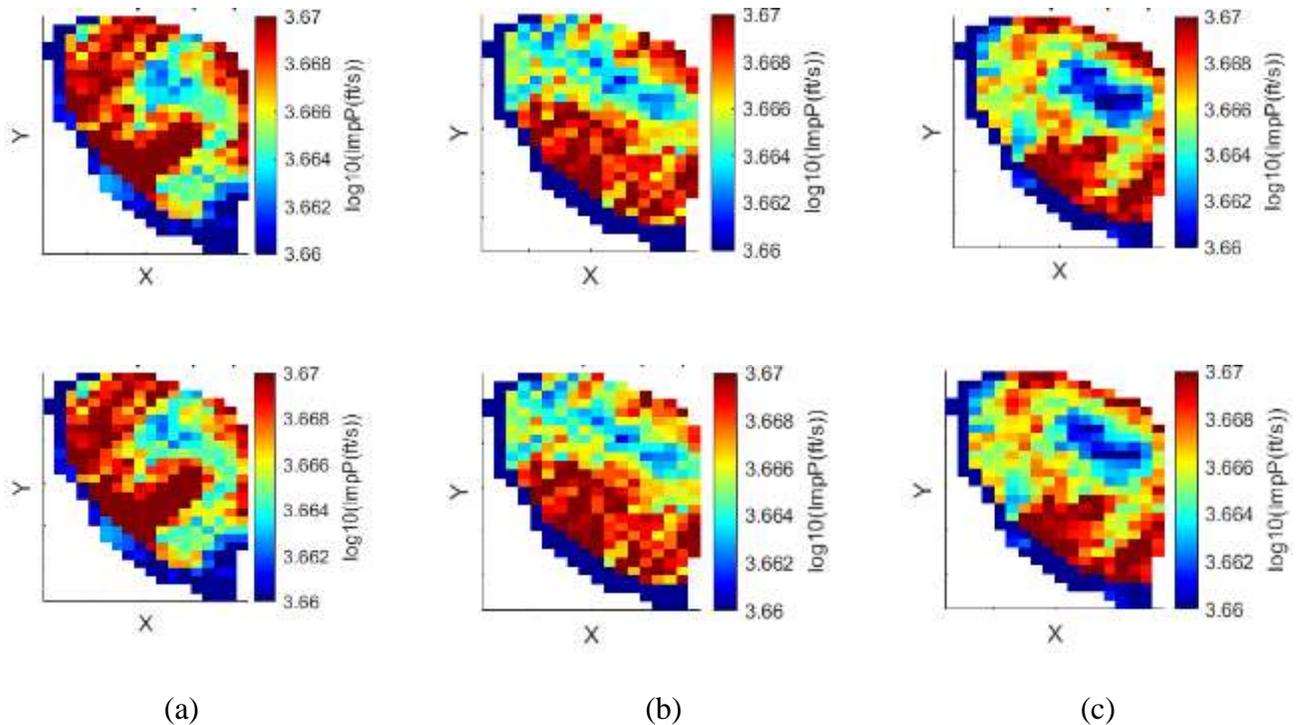

(a)                          (b)                          (c)

**Figure 6**: Acoustic impedance reconstruction showing the two survey times (first survey on-top, second survey below): (a) Impedance recovered with ES-MDA, (b) Impedance recovered with SHM-KED and (c) True impedance

Figure 6 shows the accuracy of our method in reconstructing the true impedance. SHM-KED recovered better the impedance signatures compared to standard ES-MDA with no sparse adaptation and no impedance data being assimilated. The proposed method was capable of recovering the streaks found in the true model.

Figure 7(a-c) shows the initial ensemble responses of the well bottom hole pressure, water cut and gas oil ratio. Notice the wide uncertainty in the ensemble to the true data. This is expected as we have yet to commence data assimilation. For the ES-MDA algorithm, Figures 8(a-c), we notice the uncertainty reduced but the algorithm fails to match the water cut at PRO-12. For the SHM-KED algorithm, Figures 9(a-c) we notice the algorithm does better and matches the water cut at PRO-12.

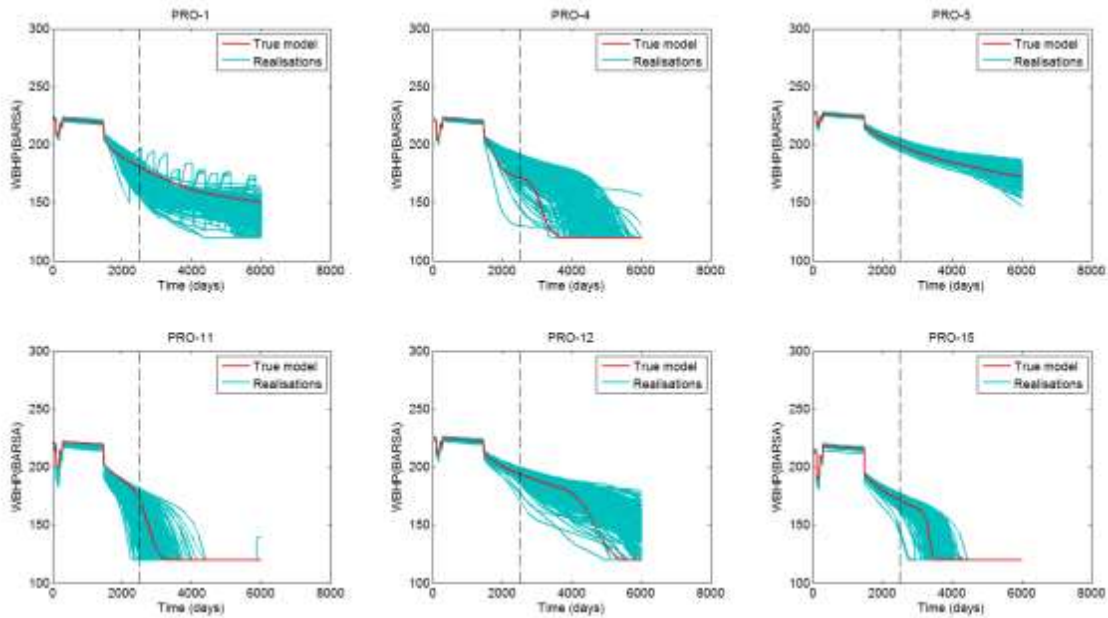

**Figure 7(a)**: shows the bottom-hole-pressure for the six producer wells of the initial ensemble. The red curve represents the true data, and the cyan overlay lines represent the realisations. The vertical dashed line represents the historical (left of this line) and prediction stages (right of this line)

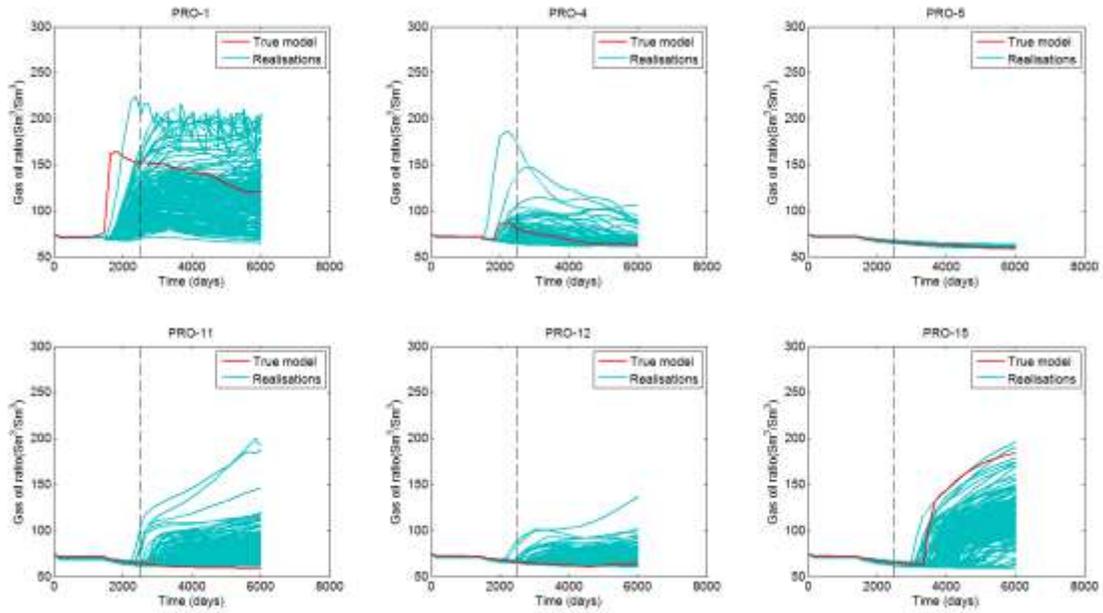

**Figure 7(b)**: shows the Gas-oil-ratio cut for the six producer wells of the initial ensemble. The red curve represents the true data, and the cyan overlay lines represent the realisations. The vertical dashed line represents the historical (left of this line) and prediction stages (right of this line)

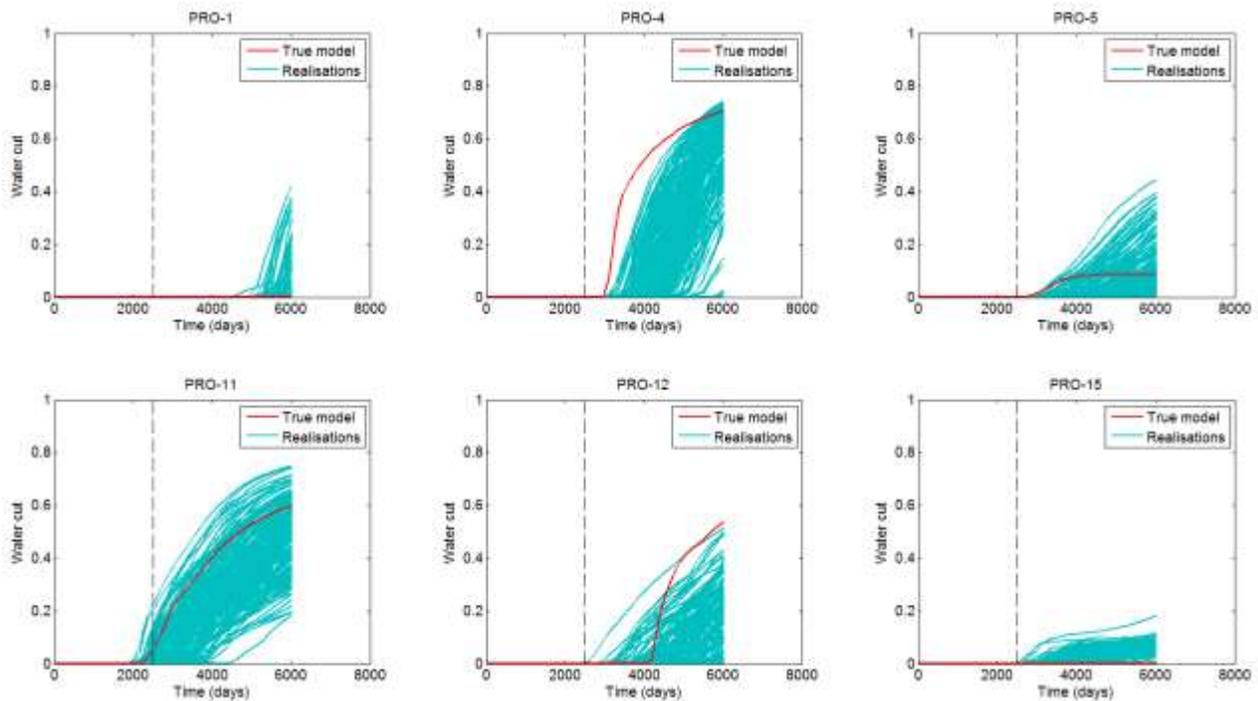

**Figure 7(c)**: shows the water cut for the six producer wells of the initial ensemble. The red curve represents the true data, and the cyan overlay lines represent the realisations. The vertical dashed line represents the historical (left of this line) and prediction stages (right of this line)

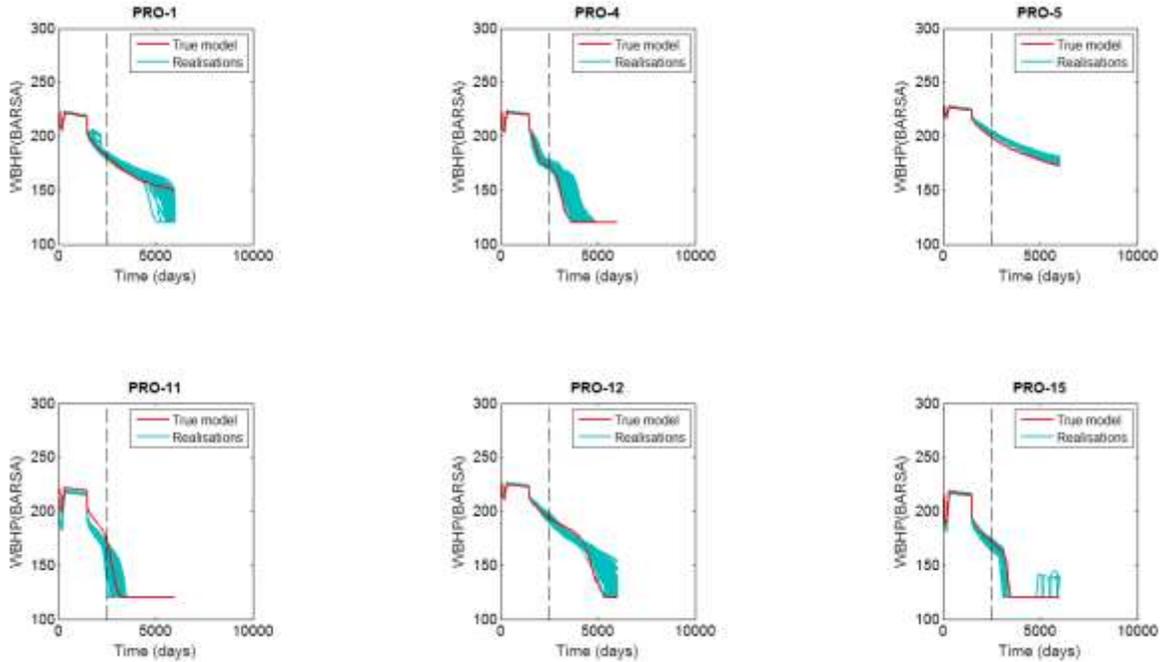

**Figure 8(a)**: shows the well bottom hole pressure for the six producer wells of the ES-MDA ensemble. The red curve represents the true data, and the cyan overlay lines represent the realisations. The vertical dashed line represents the historical (left of this line) and prediction stages (right of this line)

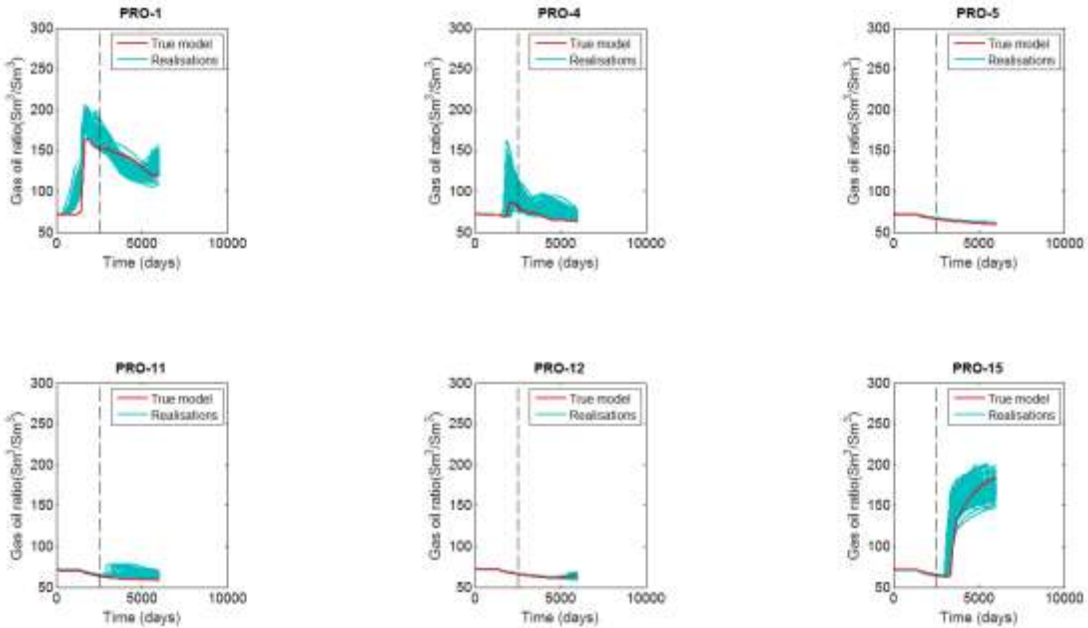

**Figure 8(b)**: shows the Gas-oil-ratio for the six producer wells of the ES-MDA ensemble. The red curve represents the true data, and the cyan overlay lines represent the realisations. The vertical dashed line represents the historical (left of this line) and prediction stages (right of this line)

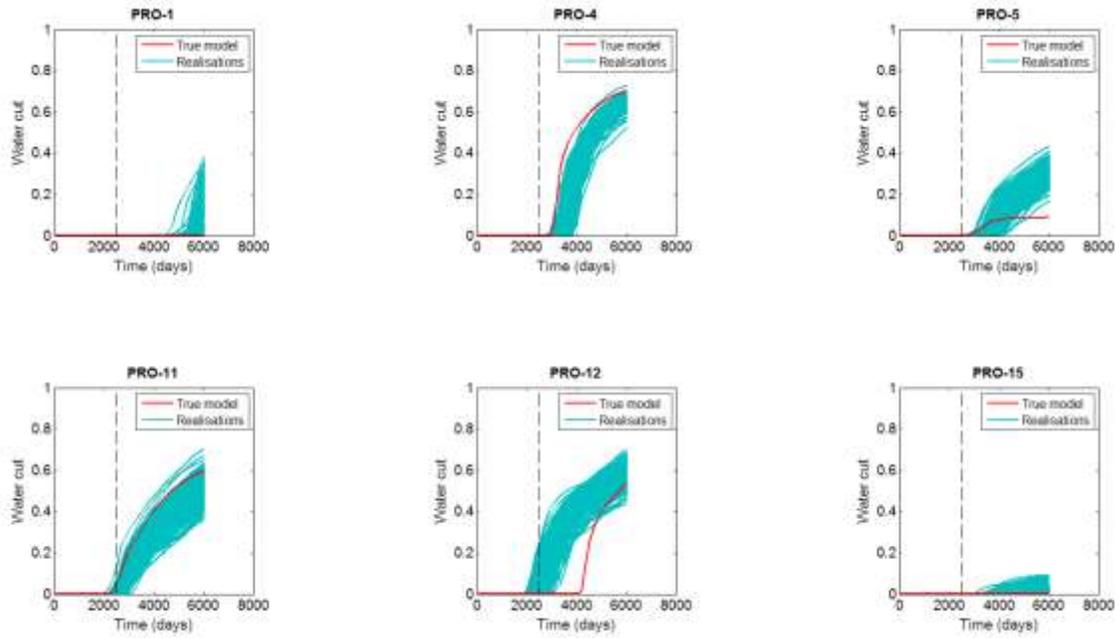

**Figure 8(c)**: shows the water cut for the six producer wells of the ES-MDA ensemble. The red curve represents the true data, and the cyan overlay lines represent the realisations. The vertical dashed line represents the historical (left of this line) and prediction stages (right of this line)

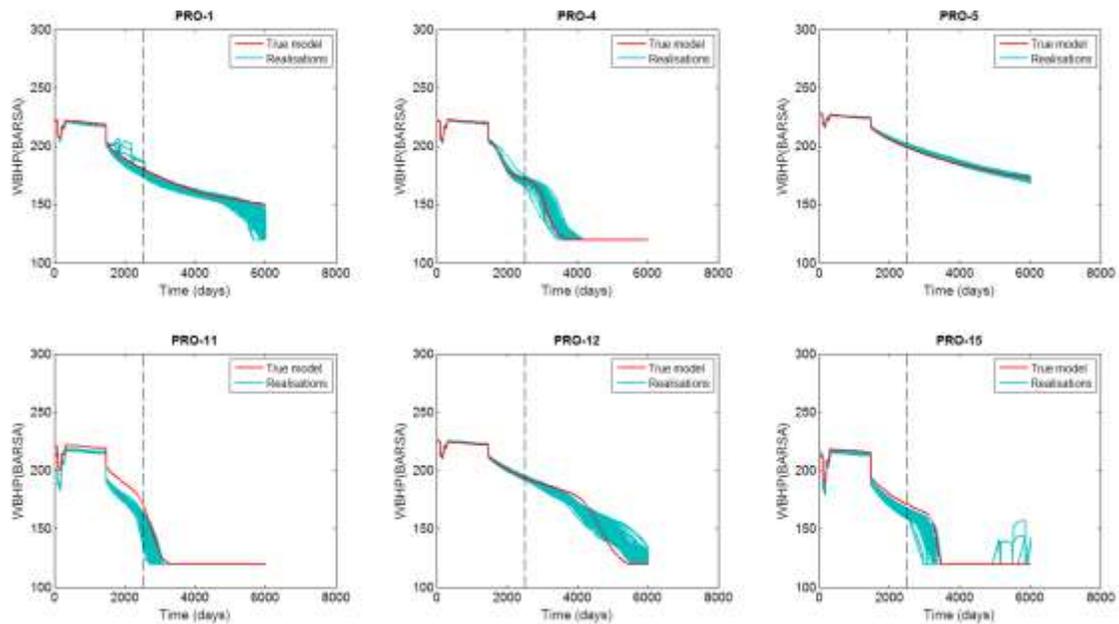

**Figure 9(a)**: shows the well bottom hole pressure for the six producer wells of the ensemble recovered by SHM-KED. The red curve represents the true data, and the cyan overlay lines represent the realisations. The vertical dashed line represents the historical (left of this line) and prediction stages (right of this line)

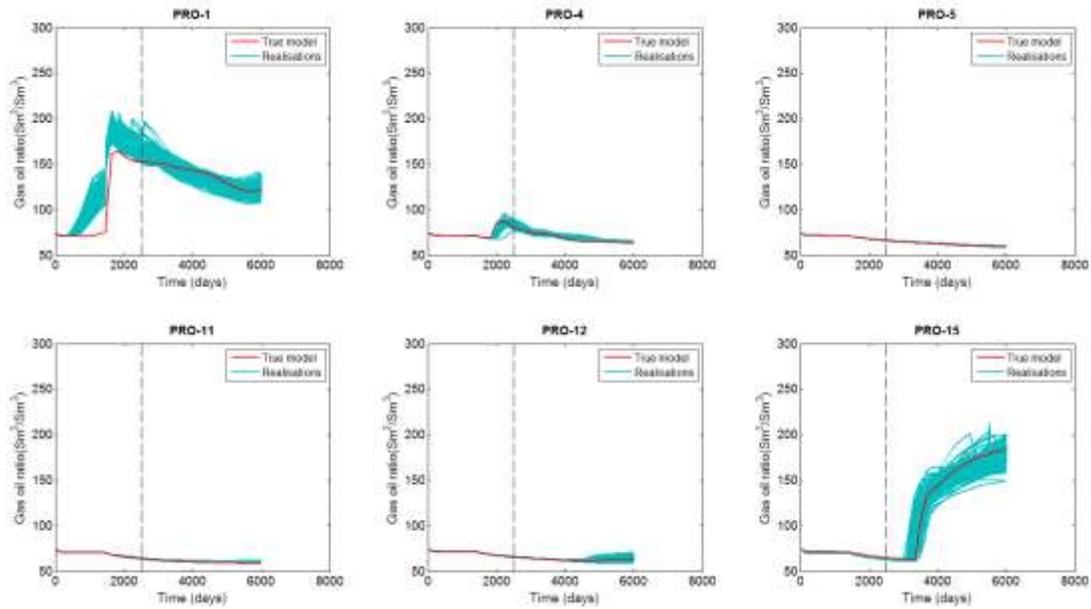

**Figure 9(b)**: shows the gas-oil-ratio for the six producer wells of the ensemble recovered by SHM-KED. The red curve represents the true data, and the cyan overlay lines represent the realisations. The vertical dashed line represents the historical (left of this line) and prediction stages (right of this line)

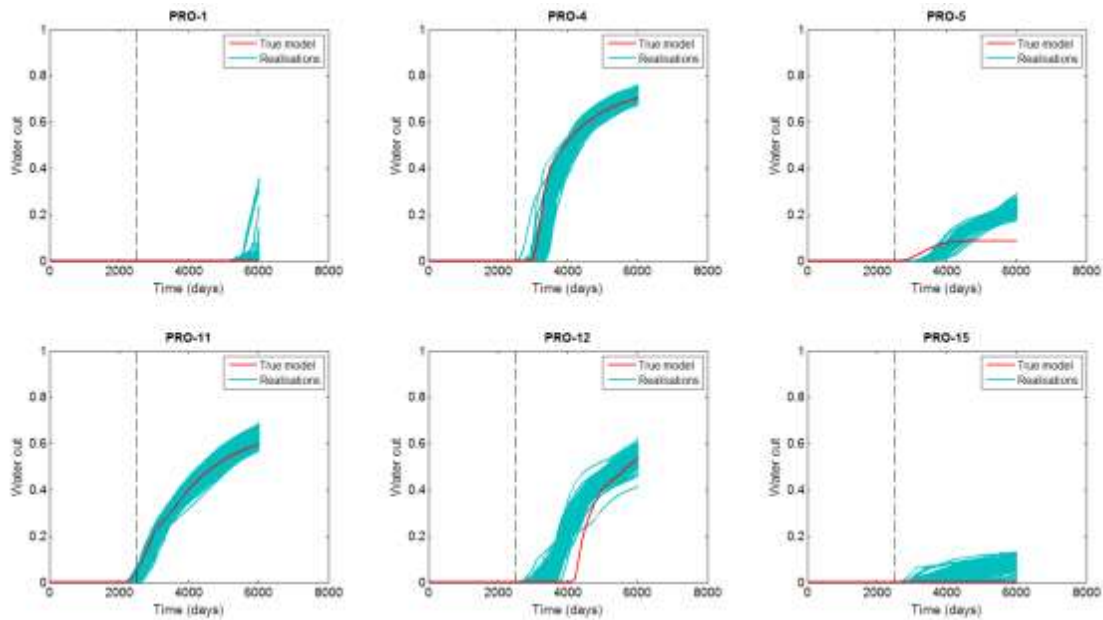

**Figure 9(c)**: shows the water cut for the six producer wells of the ensemble recovered by SHM-KED. The red curve represents the true data, and the cyan overlay lines represent the realisations. The vertical dashed line represents the historical (left of this line) and prediction stages (right of this line)

An open source algorithm **S**tructural **S**imilarity **I**ndex *(SSIM)* (Wang, et al., 2004) is used to compare the SHM-KED and ES-MDA history matched reconstructed permeability realisation to the true model. *SSIM* is an image metric quantifier that analyses the visual impact of three identities of an image: structure, contrast, and luminance ( $s$, $c$, and $l$). Explaining *SSIM* further, a value of 1 means complete similarity whiles the value of $-1$ indicates complete dissimilarity.

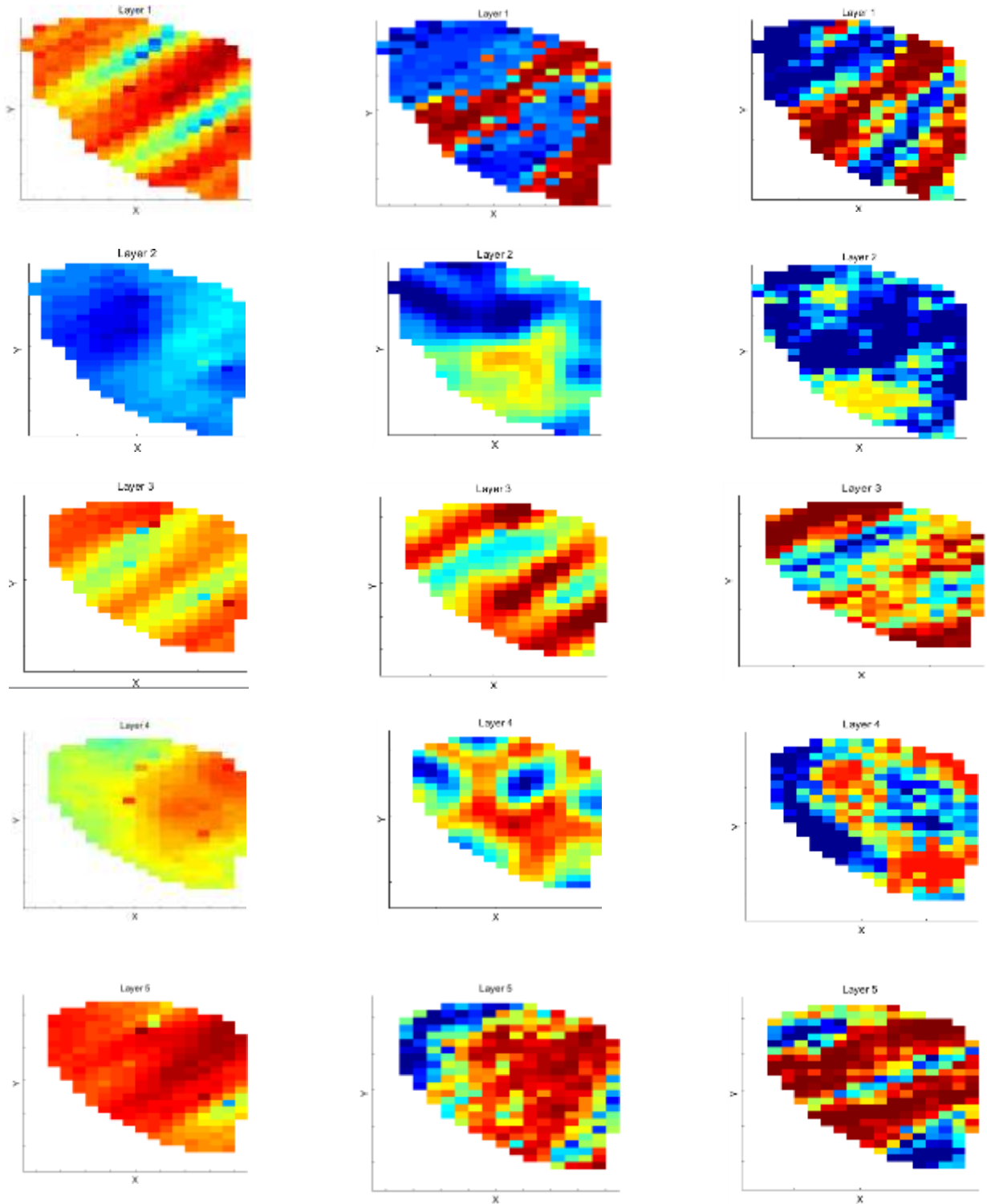

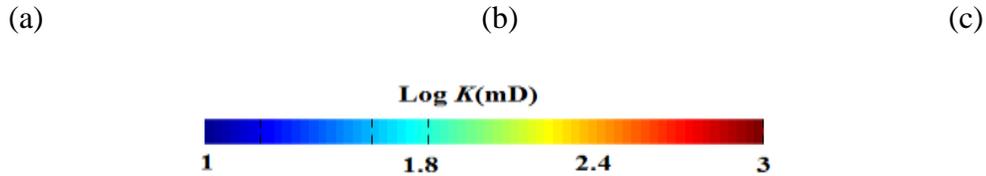

(a)                                  (b)                                  (c)

**Figure 10(a)**: shows 2D mean permeability field reconstruction

Figure 10(a) shows the 2D layer by layer mean permeability field and compares the performance of the SHM-KED with the true model directly. We observe that the algorithm is capable of reconstructing the main streak of the true model. Figure 10(b) shows the 3D analogues to Figure 10(a) with the 6 producer wells.

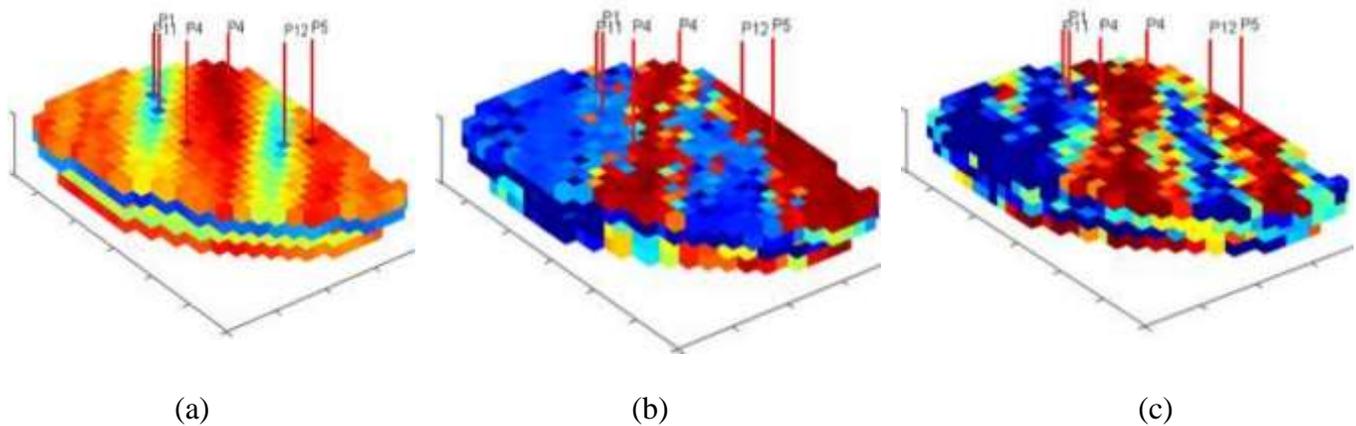

(a)                                  (b)                                  (c)

**Figure 10(b)**: 3D mean permeability field reconstruction (a) mean initial permeability field, (b) mean permeability recovered by SHM-KED and (c) True permeability field

To calculate the performance of the SHM-KED with ES-MDA, I use 2 performance indexes, namely the RMSE from Eqn 30 and *SSIM* . I then derive the combined NORM value as described in Eqn 32.

**Table 2**: RMSE values of some selected models

| Realisation no. | Initial RMSE value | Final RMSE value using ES-MDA | Final RMSE value using SHM-KED |
|---|---|---|---|
| 61 | 17.33 | 2.21 | 0.25 |
| 11 | 8.83 | 0.98 | 0.28 |
| 31 | 5.87 | 1.58 | 0.36 |
| 67 | 8.43 | 2.43 | 0.09 |

$$Combined\ NORM = \frac{(1 - SSIM) + RMSE}{2}$$

Eqn 32

**Table 3**: Combined NORM values of mean permeability fields from initial, ES-MDA and SHM-KED ensemble

| Model. | *SSIM* | *RMSE* | *Combined NORM* |
|---|---|---|---|
| Initial | 0.23 | 10.98 | 5.88 |
| ES-MDA | 0.48 | 4.58 | 2.55 |
| SHM-KED | 0.82 | 1.43 | 0.81 |

To have a complete view of how the SHM-KED outperforms ES-MDA we look at the RMSE values for the whole 200 realisations. In Figure 11(a), we see the initial RMSE of all the realisations with high values. ES-MDA (shown in Figure 11(b)) reduces these RMSE values with SHM-KED (shown in Figure 11(c)) reducing it even further.

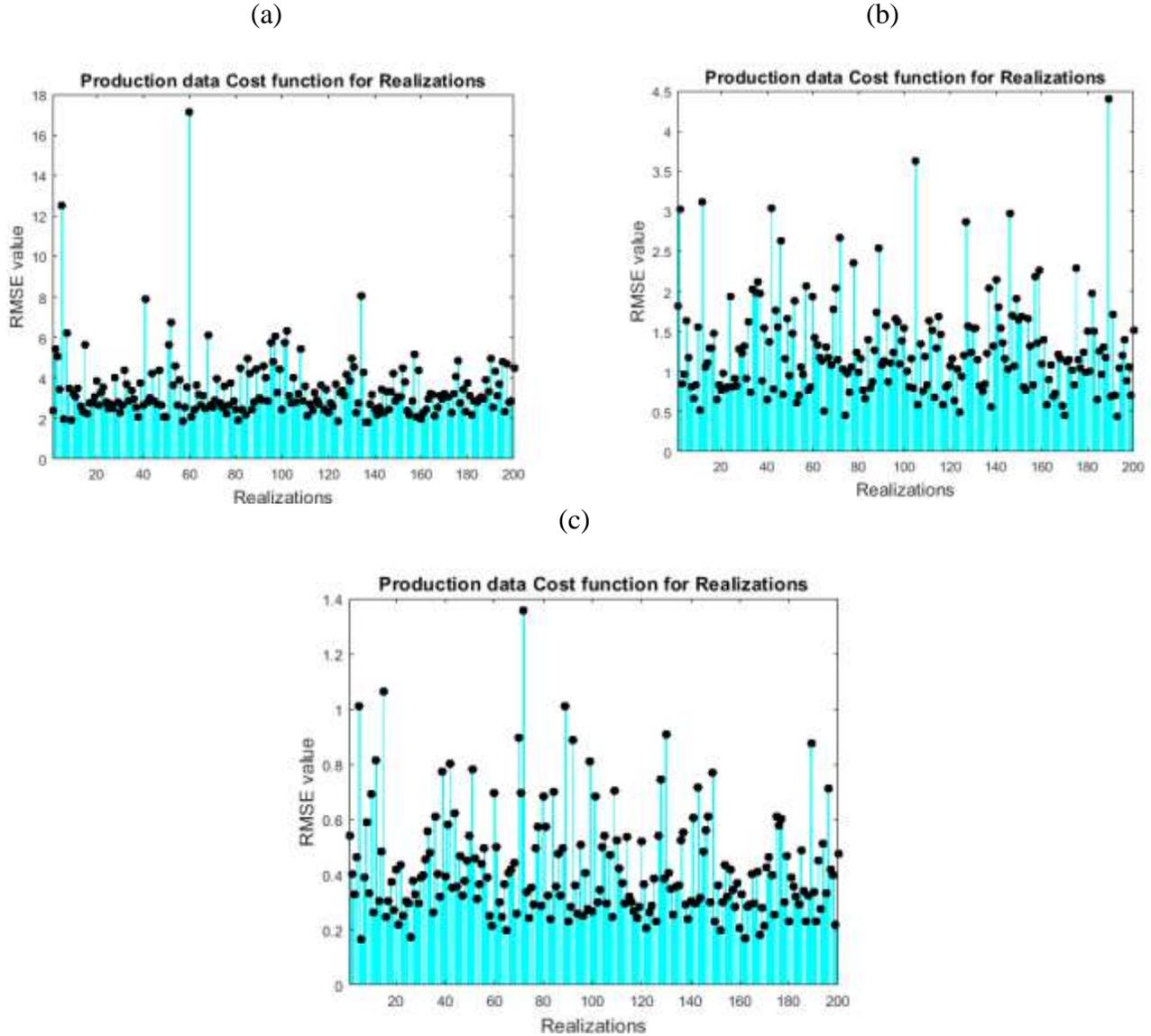

**Figure 11**: RMSE comparison (a) RMSE for initial ensemble, (b) RMSE for ensemble recovered by ES-MDA and (c) RMSE for ensemble recovered by SHM-KED

## 4. Summary and Conclusions

The SHM-KED algorithm can calibrate a non-Gaussian synthetic reservoir model by updating the sparse representations of the permeability field. The first initial library of geological permeability models was generated using *FILTERSIM* which was then used for creating the K-SVD over-complete dictionary. Dimensionality reduction of the seismic attributes, in particular, acoustic impedance, constructed using DCT is effective in reducing the size of the acoustic impedance library by selecting leading coefficients concerning the DCT transform basis. The permeability field is converted to sparse coefficients using this K-SVD dictionary together with OMP. The weight matrix obtained by decomposing the permeability field is imported to ES-MDA as a state vector. In each assimilation of ES-MDA, an update of weights results in reservoir models that are well conditioned to both static and dynamic data.

The main novelty of this paper is the incorporation of sparse representations of permeability and acoustic impedance updates which is practically achieved by using a combination of machine learning and ensemble-based data assimilation methods. History-matching results of the PUNQ-S3 synthetic reservoirs indicate that the developed method is capable in capturing the permeability streak of the true reservoir model and the same time reconstructing the acoustic impedance of the true model. We also observe an improved matching accuracy in both history and forecast in terms of well production, the reduced dispersion of production behaviours and permeability distribution, and the well-connected channel body of reservoir models with geological plausibility. ES-MDA with the dictionary update and sparse representation of the impedance yield higher matching accuracy values and lower dispersion values than ES-MDA incorporated without any sparse parametrisation. The increase in computational costs invested during the initial construction of the dictionary (done once and off-line) is affordable compared to the assimilation algorithms not coupled with sparse coding. Improving the matching accuracy of some producer wells remains an outstanding task for the proposed technique despite the overall enhanced matching quality. Future works will be to adopt an adaptive inflation factor during the ES-MDA data assimilation and a surrogate based modelling for the forwarding of reservoir production data to reduce the expensive forwarding sequence using a full fidelity run. Also, a supervised learning technique involving deep neural networks for constructing these surrogate models or for permeability parametrisation is also planned for future research.

## Acknowledgement

This project has been supported by the Niger Delta Development Commission of the Federal Republic of Nigeria (NDDC) and the University of Manchester. Thanks are due to Schlumberger for providing the reservoir simulator.